\documentclass[journal]{IEEEtran}

\ifCLASSINFOpdf
\else
   \usepackage[dvips]{graphicx}
\fi
\usepackage{url}
\usepackage{booktabs}
\usepackage{amsmath,amsfonts}
\usepackage{graphicx}
\usepackage{url}
\usepackage{multirow} 
\usepackage{graphicx}

\usepackage{xcolor}
\usepackage{soul}
\usepackage[normalem]{ulem}

\soulregister\cite7
\soulregister\ref7
\soulregister\pageref7

\begin{document}

\title{FADRW: A Feature-Aware Modulated and Dynamically Reweighted Loss for Few-Shot Linguistic Steganalysis}

\author{Shuo~Liu,  
        Xianghong~Lin, 
        Yukun~Wei, 
        and~Zhongliang~Yang
\thanks{S. Liu is with the International School, Beijing University of Posts and Telecommunications, Beijing 100876, China. X. Lin, Y. Wei, and Z. Yang are with the School of Cyberspace Security, Beijing University of Posts and Telecommunications, Beijing 100876, China (e-mail: yangzl@bupt.edu.cn).}%
\thanks{(*Corresponding author: Zhongliang Yang.)}%
}

\markboth{IEEE Signal Processing Letters,~Vol.~XX, No.~X, Month~2026}%
{Liu \MakeLowercase{\textit{et al.}}: FADRW for Few-Shot Linguistic Steganalysis}
\maketitle

\begin{abstract}
The ubiquity of social media platforms facilitates malicious linguistic steganography, posing significant security risks. However, detection is severely hampered by two fundamental issues during model training. Firstly, extreme class imbalance (less than 1\% steganographic samples) induces a strong decision bias. Secondly, the invisibility of generative steganography means its features are nearly indistinguishable from benign text; this similarity, compounded by their extreme rarity, leads to severe feature marginalization, where faint steganographic signals are completely overwhelmed. To directly address these optimization-level challenges, we propose FADRW (Feature-Aware Modulated and Dynamically Reweighted Loss), a novel loss function framework engineered for few-shot steganalysis. FADRW employs Dynamic Reweighting to progressively counteract decision bias, and a Feature-Aware Modulation module to structurally reshape the feature space, preventing feature marginalization by enhancing the separability of these subtle features. Extensive experiments on datasets from three real-world social platforms demonstrate that FADRW significantly outperforms state-of-the-art methods, particularly in the challenging few-shot steganographic sample scenario.
\end{abstract}


\IEEEpeerreviewmaketitle

\section{Introduction}

\IEEEPARstart{S}{teganography} aims to hide secret information in seemingly ordinary carriers so that the existence of communication remains concealed \cite{cox2007digital}. Compared with encryption, which protects content while exposing the presence of a message, steganography attempts to make steganographic carriers indistinguishable from normal carriers. This property makes it attractive not only for privacy-preserving communication, but also for malicious uses such as covert coordination, data exfiltration, and cyberattacks \cite{bieniasz2018towards}. Therefore, steganalysis, which aims to detect the presence of hidden payloads within seemingly benign cover media, is of great research significance for maintaining cyberspace security.

Among various media, text is a particularly attractive carrier for steganography because it is ubiquitous, easy to disseminate, and remains stable during online communication. The rapid growth of social media further amplifies this advantage by providing massive volumes of short, interactive, and highly diverse text, thereby creating favorable environments for covert communication. Recent linguistic steganography has evolved from early neural generative methods, such as RNN-Stega and VAE-Stega, to more advanced neural and LLM-based paradigms \cite{yang2018rnn,ziegler2019neural,yang2020vae,lin2024zeroshot,wu2024llmstega}. These methods embed secret information by modulating token selection during text generation, improving both linguistic fluency and payload flexibility. More recent studies have further moved toward provably secure linguistic steganography
\cite{ding2023discop,liao2025framework,bai2025shimmer,wang2025sparsamp,pan2025prefix},
where the generated stegotext is designed to better match the
distribution of benign text. Consequently, modern stegotext becomes increasingly natural and statistically concealed, making reliable detection substantially more challenging.

To detect such hidden communication, steganalysis models aim to capture subtle discrepancies between steganographic and normal texts. Early studies mainly relied on handcrafted statistical cues, such as lexical or synonym-based features \cite{chen2011steganalysis,xiang2014linguistic}. With the development of neural methods, deep models such as TS-RNN \cite{yang2019ts}, TS-CSW \cite{yang2020ts}, FastText-SM \cite{yang2019fast}, TS-GNN \cite{wu2021linguistic}, and HSL-DA \cite{peng2023text} significantly improved single-text steganalysis by learning more discriminative representations directly from text. However, isolated text analysis is often insufficient for social media, where messages are short, fragmented, and semantically sparse. To alleviate this issue, recent studies started to incorporate conversational and relational context. LSTSN \cite{yang2022linguistic} showed that surrounding context can substantially improve detection in social networks. Subsequent works further exploited interaction structures, external knowledge, and graph-based modeling, such as LINK~\cite{yang2023link}, CATS~\cite{pang2023cats}, TGCA~\cite{lu2025tgca}, STLC-KG~\cite{wang2025stlc}, CASS-LS~\cite{jiang2025cassls}, UP4LS~\cite{wang2026up4ls} and ATS-MPFF~\cite{xu2026atsmpff}, achieving stronger context-aware steganalysis performance.

Despite this progress, a critical challenge remains insufficiently addressed: extreme class imbalance in real-world social platforms. In practice, steganographic texts are exceedingly rare, often accounting for less than 1\% of all messages, which makes social media steganalysis essentially a few-shot detection problem. Under this setting, two optimization-level difficulties arise. First, the subtle features of steganographic texts are easily submerged by the overwhelming number of benign samples, leading to weak discriminability in the learned feature space. Second, the classifier is dominated by the majority class during training, which biases the decision boundary toward normal texts and weakens generalization to rare steganographic samples. Therefore, to build a robust steganalysis model for real-world imbalanced data, it is critical to tackle the problems of feature marginalization and decision bias at the optimization level.

\begin{figure*}[t]
\centering
\includegraphics[width=0.9\textwidth]{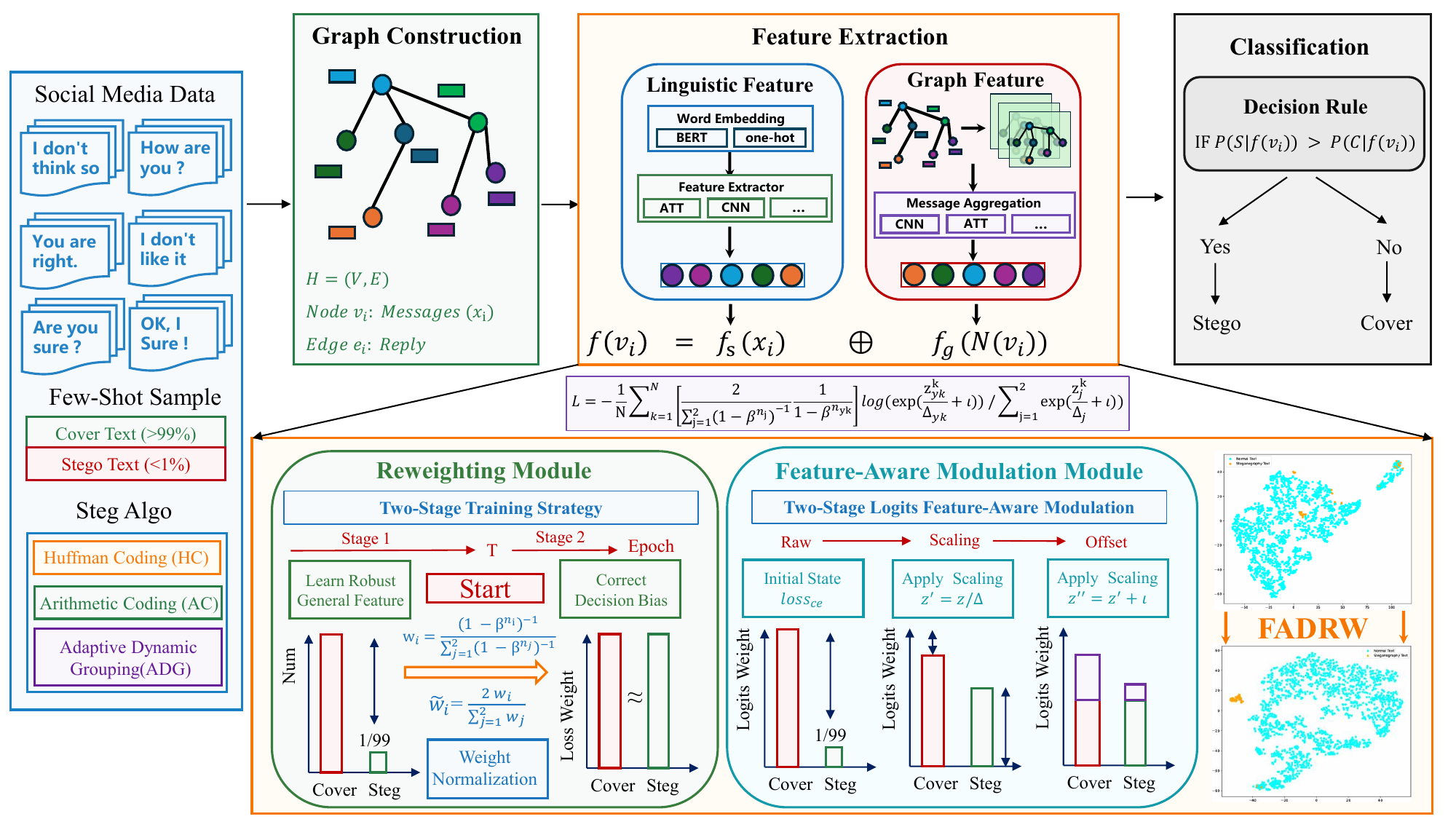}
\caption{The FADRW framework for few-shot steganalysis. After constructing a graph from imbalanced social media data and extracting features, the model is optimized by our FADRW loss. Its Reweighting Module corrects decision bias through staged training, while its Feature-Aware Modulation Module prevents feature marginalization. The t-SNE visualization on the right confirms FADRW's effectiveness in creating a more separable feature space.}
\label{fig:fig2}
\end{figure*}

To address these challenges about few-sample steganalysis, our main contributions are threefold:
\begin{itemize}
    \item We identify decision bias, arising from extreme class imbalance, as a primary failure mode in few-shot steganalysis. To counteract this, we propose a Dynamic Reweighting strategy, which utilizes a staged training process to progressively correct the model’s focus and refine its decision boundary.
    \item We identify feature marginalization, where the inherently subtle and rare features of steganographic text are overwhelmed, as the second key bottleneck. We introduce a novel Feature-Aware Modulation module that structurally reshapes the feature space by scaling and shifting logits, thereby enhancing the separability of these faint but critical signals.
    \item We conduct extensive experiments on datasets from three social platforms (Reddit, X, and Weibo). The results demonstrate that our FADRW Loss, which synergistically combines both contributions, significantly outperforms state-of-the-art methods, particularly in the most challenging less than 1\% steganographic sample scenarios.
\end{itemize}

\section{Method}
\subsection{Task Modeling and FADRW Overview}
We model the social media conversation as a graph $H=(V,E)$, where $V=\{v_1,v_2,\ldots,v_n\}$ denotes the set of message nodes and $E\subseteq V\times V$ denotes the set of reply edges, with $(v_i,v_j)\in E$ iff $v_j$ is a direct reply to $v_i$. Each node $v_i\in V$ corresponds to a message text $x_i$ and is associated with a binary label $y_i\in\{0,1\}$, where $y_i=1$ denotes a steganographic text and $y_i=0$ denotes a normal cover text. The goal is to predict the label of each node by jointly exploiting its textual content and conversational context.

As illustrated in Fig.~1, the feature extraction stage contains two parallel branches. In the linguistic branch, $x_i$ is mapped into word-level embeddings and encoded by an attention-enhanced CNN(ATT-CNN) to obtain the linguistic feature $f_s(x_i)$. In the graph branch, the neighborhood information of $v_i$ is aggregated through a GNN-based message-aggregation module along the reply structure to produce the graph contextual feature $f_g(\mathcal{N}(v_i))$. The final node representation is obtained by feature fusion:
\begin{equation}
f(v_i)=f_s(x_i)\oplus f_g(\mathcal{N}(v_i)),
\end{equation}
where $\mathcal{N}(v_i)$ denotes the set of neighboring nodes of $v_i$, and $\oplus$ denotes concatenation along the feature dimension.

However, directly optimizing the classifier on $f(v_i)$ is severely
hindered by extreme class imbalance, which causes majority-induced
decision bias and feature marginalization of subtle steganographic
signals. To address these challenges, we propose FADRW, an
optimization-level loss framework consisting of Dynamic Reweighting and
Feature-Aware Modulation, which respectively alleviate classifier bias
and enhance the separability of weak steganographic features. The
detailed formulations are presented in the following subsections.

\subsection{Dynamic Reweighting for Decision Bias}
To counteract the decision bias in steganalysis, where the optimization is dominated by $>$ 99\% benign samples, pushing the decision boundary to misclassify rare stego examples, FADRW employs a Dynamic Reweighting strategy. Our strategy mitigates this with a progressive, staged training scheme, rather than a fixed factor, allowing the model to first learn robust general features before shifting focus to the minority class. The reweighting intensity $\beta$ is selected from a predefined schedule $(\beta_{0},\beta_{1},...)$ based on the current training stage:
\begin{equation}
\beta=\beta_{k} \quad \text{where} \quad k=\lfloor\frac{\text{epoch}}{T}\rfloor,
\end{equation}
where $T$ is the length of each training stage. This typically involves an initial stage with no reweighting (e.g., $\beta_{0}=0$) to learn robust initial features, followed by later stages with aggressive reweighting (e.g., $\beta_{1}=0.999$) to correct the bias. Using the selected $\beta$, we then calculate the final normalized weight $\tilde{w}_{i}$ for class $i$. This weight is inversely proportional to the class's effective number of samples, ensuring the minority steganographic class receives higher emphasis. For the binary steganalysis task, the weight is:
\begin{equation}
\bar{w}_{i} = \frac{2(1 - \beta^{n_i})^{-1}}{\sum_{j=0}^{1} (1 - \beta^{n_j})^{-1}},
\end{equation}
where $n_{i}$ is the number of training samples in class $i$. By applying these progressively increasing weights, we force the model to refine its decision boundary, ensuring that the few, yet critical, steganographic samples contribute substantially to the total loss in the final training stages.

\subsection{Feature-Aware Modulation for Feature Marginalization}
To combat the feature marginalization endemic to imbalanced steganalysis, we introduce a Logits-based Feature-Aware Modulation module. Under a standard Cross-Entropy (CE) loss, the subtle, low-magnitude features characteristic of steganographic texts are overwhelmed by the gradients from the vast number of benign samples. Our module prevents this by structurally reshaping the feature space before the final loss calculation. This is achieved through a two-fold modulation of the model's output logits. First, a class-aware scaling factor $\Delta_{i}$ enforces a larger decision margin for the minority class. For our binary task, it is defined as:
\begin{equation}
\Delta_{i}=\frac{n_{i}^{-\gamma}}{\min_{j\in\{0,1\}}n_{j}^{-\gamma}},
\end{equation}
where $n_{i}$ is the sample count for class $i$, and the hyperparameter $\gamma$ controls the scaling intensity. Second, this is complemented by an additive shifting offset $\iota_{i}$ that directly counterbalances the inherent class priors. By incorporating the class frequency directly, the offset is calculated as:
\begin{equation}
\iota_{i}=\tau \ln \left(\frac{n_{i}}{\sum_{j=0}^{1}n_{j}}\right),
\end{equation}
where the hyperparameter $\tau$ controls the shifting strength. These modulations act more strongly on the minority class, where a small sample count leads to a larger scaling factor and a stronger offset. As a result, the adjusted logits enlarge minority-class separability and compensate for its disadvantaged prior. These two components are then integrated to produce the final modulated logits, which are used to compute the loss. The complete modulated logit for class $i$ of a sample $k$ is given by:

\begin{table*}[t]
\centering
\caption{Positive-class F1 scores (\%, 5-run average) on the Reddit, Sina, and Tweet datasets under various steganographic algorithms and embedding payloads (BPW).}
\label{tab:table1}
\begin{tabular}{l l ccccc ccc c}
\toprule
\multirow{2}{*}{\textbf{Dataset}} & \multirow{2}{*}{\textbf{Algorithm}} & \multicolumn{5}{c}{\textbf{AC (BPW)}} & \multicolumn{3}{c}{\textbf{HC (BPW)}} & \textbf{ADG (BPW)} \\
\cmidrule(lr){3-7} \cmidrule(lr){8-10} \cmidrule(lr){11-11}
 & & \textbf{0.24} & \textbf{1.17} & \textbf{2.00} & \textbf{2.66} & \textbf{3.27} & \textbf{1.86} & \textbf{2.59} & \textbf{3.23} & \textbf{4.13} \\
\midrule
\multirow{5}{*}{\textbf{Reddit}} 
 & LSTSN \cite{yang2022linguistic} & 89.58 & \textbf{87.46} & 82.68 & 78.22 & 68.45 & 83.98 & 84.58 & 76.83 & 3.45 \\
 & LDAM \cite{cao2019learning} & 89.25 & 87.29 & 82.70 & 78.67 & 69.29 & 85.08 & 83.86 & 75.66 & 16.92 \\
 & Focal \cite{lin2017focal} & 89.64 & 87.45 & \textbf{84.19} & 76.74 & 65.03 & 84.39 & 83.61 & 76.04 & 0.93 \\
 & CE+SAM \cite{foret2020sharpness} & 88.75 & 86.92 & 83.34 & 77.44 & 65.63 & 84.18 & 81.73 & 74.94 & \textbf{20.64} \\
 & \textbf{FADRW(Proposed)} & \textbf{89.97} & 87.40 & 83.56 & \textbf{79.19} & \textbf{73.50} & \textbf{85.83} & \textbf{85.02} & \textbf{79.42} & 12.43 \\
\midrule
\multirow{5}{*}{\textbf{Sina}} 
 & LSTSN \cite{yang2022linguistic} & 73.94 & 52.07 & 41.03 & 17.98 & 20.87 & 57.17 & 21.05 & 23.65 & 4.66 \\
 & LDAM \cite{cao2019learning} & 73.29 & 52.69 & 39.79 & 18.21 & 16.62 & 57.44 & 27.86 & 25.05 & 14.54 \\
 & Focal \cite{lin2017focal} & 74.24 & 53.75 & 31.69 & 16.96 & 11.55 & 56.04 & 12.63 & 17.86 & 6.41 \\
 & CE+SAM \cite{foret2020sharpness} & 70.10 & 51.70 & 45.88 & 32.51 & 30.37 & 60.47 & 38.44 & 34.88 & \textbf{23.30} \\
 & \textbf{FADRW(Proposed)} & \textbf{77.27} & \textbf{62.22} & \textbf{50.37} & \textbf{36.48} & \textbf{35.15} & \textbf{67.16} & \textbf{42.93} & \textbf{43.96} & 22.67 \\
\midrule
\multirow{5}{*}{\textbf{Tweet}} 
 & LSTSN \cite{yang2022linguistic} & 64.33 & 64.90 & 67.01 & 66.85 & 58.69 & 66.25 & 69.44 & 67.48 & 0.95 \\
 & LDAM \cite{cao2019learning} & 67.52 & 67.83 & 66.29 & 66.61 & 59.90 & 67.84 & 68.60 & 66.21 & 17.31 \\
 & Focal \cite{lin2017focal} & 63.49 & 65.83 & 67.80 & 66.88 & 60.20 & 67.58 & 68.33 & 65.24 & 9.96 \\
 & CE+SAM \cite{foret2020sharpness} & 70.83 & 69.37 & 67.68 & \textbf{69.07} & 60.35 & 68.51 & 70.05 & 67.33 & \textbf{19.34} \\
 & \textbf{FADRW(Proposed)} & \textbf{73.50} & \textbf{71.35} & \textbf{68.47} & 68.06 & \textbf{63.73} & \textbf{69.99} & \textbf{70.15} & \textbf{68.31} & 15.69 \\
\bottomrule
\end{tabular}
\end{table*}

\begin{table*}[t]
\centering
\caption{Ablation study of FADRW components on the Reddit, Sina, and Tweet datasets. Positive-class F1 (\%, 5-run average) is reported for the full model (FADRW), w/o Dynamic Reweighting, and w/o Feature-Aware Modulation.}
\label{tab:table2}
\begin{tabular}{l l ccccc ccc c}
\toprule 
\multirow{2}{*}{\textbf{Dataset}} & \multirow{2}{*}{\textbf{Algorithm}} & \multicolumn{5}{c}{\textbf{AC (BPW)}} & \multicolumn{3}{c}{\textbf{HC (BPW)}} & \textbf{ADG (BPW)} \\
\cmidrule(lr){3-7} \cmidrule(lr){8-10} \cmidrule(lr){11-11} 
 & & \textbf{3.27} & \textbf{3.76} & \textbf{4.20} & \textbf{4.52} & \textbf{4.88} & \textbf{1.86} & \textbf{3.23} & \textbf{3.82} & \textbf{4.13} \\
\midrule 
\multirow{4}{*}{\textbf{Reddit}} 
 & \textbf{FADRW (Full)} & \textbf{73.50} & 67.97 & \textbf{62.38} & \textbf{55.21} & \textbf{44.17} & \textbf{85.83} & \textbf{79.42} & \textbf{77.06} & 12.43 \\
 & \textbf{w/o Feature-Aware} & 68.56 & 64.28 & 45.98 & 35.26 & 15.78 & 84.56 & 76.25 & 71.16 & 8.76 \\
 & \textbf{w/o Reweighting} & 72.81 & \textbf{69.08} & 54.35 & 54.94 & 25.30 & 84.37 & 77.60 & 75.24 & \textbf{16.92} \\
\midrule
\multirow{4}{*}{\textbf{Sina}} 
 & \textbf{FADRW (Full)} & \textbf{35.15} & \textbf{26.16} & \textbf{26.71} & \textbf{18.90} & 19.48 & 67.16 & \textbf{43.96} & \textbf{36.44} & \textbf{22.67} \\
 & \textbf{w/o Feature-Aware} & 18.09 & 7.79 & 8.56 & 5.91 & 5.98 & 60.83 & 25.99 & 22.61 & 7.01 \\
 & \textbf{w/o Reweighting} & 33.69 & 26.13 & 22.44 & 17.52 & \textbf{21.68} & \textbf{69.28} & 42.93 & 34.62 & 15.49 \\
\midrule
\multirow{4}{*}{\textbf{Tweet}} 
 & \textbf{FADRW (Full)} & \textbf{63.73} & \textbf{67.71} & \textbf{64.52} & \textbf{52.20} & \textbf{47.32} & \textbf{69.99} & \textbf{68.31} & \textbf{66.84} & \textbf{15.69} \\
 & \textbf{w/o Feature-Aware} & 58.76 & 66.20 & 61.01 & 41.06 & 38.52 & 67.28 & 65.35 & 66.68 & 3.80 \\
 & \textbf{w/o Reweighting} & 60.75 & 65.94 & 59.76 & 49.69 & 43.43 & 69.95 & 67.66 & 66.02 & 15.33 \\
\bottomrule 
\end{tabular}
\end{table*}
\begin{equation}
\tilde{z}_{i}^{(k)}=\frac{z_{i}^{(k)}}{\Delta_{i}}+\iota_{i},
\end{equation}
This allows us to formulate the loss for the feature modulation component as the cross-entropy over these adjusted logits:
\begin{equation}
\mathcal{L}_{mod}=-\frac{1}{N}\sum_{k=1}^{N}\log\frac{\exp(\tilde{z}_{y_{k}}^{(k)})}{\sum_{l=0}^{1}\exp(\tilde{z}_{l}^{(k)})},
\end{equation}
where $N$ is the number of samples in the batch, and $y_{k}$ is the ground-truth label for the $k$-th sample. By substituting the standard logits with these structurally adjusted ones, our method ensures that the features of rare steganographic samples are protected from marginalization and contribute meaningfully to the optimization process, a property not achievable with standard CE loss.
By combining the dynamic reweighting term with the modulated logits, the final FADRW loss is formulated as
\begin{equation}
\begin{split}
\mathcal{L} &= - \frac{1}{N} \sum_{k=1}^{N} \left[ \frac{2}{\sum_{j=0}^{1} (1 - \beta^{n_j})^{-1}} \frac{1}{1 - \beta^{n_{y_k}}} \right] \\
&\quad \times \log \left( \frac{\exp \left( \frac{z_{y_k}^{(k)}}{\Delta_{y_k}} + \iota_{y_k} \right)}{\sum_{l=0}^{1} \exp \left( \frac{z_l^{(k)}}{\Delta_l} + \iota_l \right)} \right),
\end{split}
\end{equation}
where the reweighting term alleviates majority-induced decision bias, while the modulation term improves the separability of weak steganographic features.

\section{Experiments and Analysis}
\subsection{Data Preparation and Experimental Settings}
We evaluate the proposed method on the Stego-Sandbox dataset \cite{yang2022linguistic}, which
contains real-world conversational data collected from three social media platforms, namely Reddit, X (dataset name Tweet in \cite{yang2022linguistic}), and Weibo (dataset name Sina in \cite{yang2022linguistic}). The dataset is generated using three representative generative linguistic steganographic algorithms, namely ``Huffman Coding'' (HC) \cite{yang2018huffman}, ``Arithmetic Coding'' (AC) \cite{ziegler2019neural}, and ``Adaptive Dynamic Grouping'' (ADG) \cite{zhang2021provably}, with embedding payloads controlled by bits per word (BPW), i.e., the average number of hidden bits per word. Following the original protocol, the dataset is partitioned into training, validation, and test sets with a ratio of 75\% / 12.5\% / 12.5\%. To simulate realistic social media scenarios where steganographic content is extremely rare, we further construct highly imbalanced subsets by random sampling while preserving the original conversational graph structures, such that steganographic samples account for less than 1\% of the training data.

For a fair comparison, all methods are implemented on the same architectural backbone, LSTSN \cite{yang2022linguistic}. To evaluate the proposed loss, we compare it with imbalance-aware baselines, including LDAM~\cite{cao2019learning} and Focal Loss~\cite{lin2017focal}. We further introduce SAM~\cite{foret2020sharpness} as an optimization baseline and combine it with CE, LDAM, and Focal Loss for comparison. Since the task is highly imbalanced, we report the positive-class F1-score for the steganographic class as the main evaluation metric. All models are trained for 200 epochs, and all reported results are averaged over five independent runs. For our proposed FADRW, the optimal hyperparameters are empirically set to $\gamma=0.05$, $\tau=0.5$, and $T=20$ based on the validation set. 

\subsection{Evaluation Results and Discussion}
The comparative results in Table~\ref{tab:table1} demonstrate the effectiveness of FADRW under the less-than-1\% setting. FADRW outperforms the baseline methods in the majority of settings. It achieves the best performance in most configurations on the Reddit and Sina datasets, showing strong effectiveness under extreme class imbalance. The gains are particularly evident on the HC and AC generators, where the proposed loss is better able to preserve weak minority features and reduce the bias toward the dominant normal class. Tables~\ref{tab:table1} and~\ref{tab:table2} cover complementary BPW settings: Table~\ref{tab:table1} follows the configurations commonly used by the backbone method~\cite{yang2022linguistic} for direct comparison, while Table~\ref{tab:table2} adds BPW settings for ablation analysis to examine each FADRW component under different payloads.

The advantage of FADRW is more evident in low-payload settings, where steganographic signals are weaker. For example, on Sina with AC at 1.17 BPW, FADRW achieves an F1-score of 62.22, outperforming the baselines and showing its effectiveness in detecting subtle hidden signals.

The ablation results in Table~\ref{tab:table2} validate the effectiveness of both FADRW components. Dynamic Reweighting and Feature-Aware Modulation improve performance in most cases, and the full model generally achieves the best overall results, confirming their complementarity. A few local reversals occur when one ablated variant slightly exceeds the full model, suggesting that the dominance of decision bias and feature marginalization varies across settings. For example, on Reddit under ADG, w/o Reweighting outperforms full FADRW (16.92 vs. 12.43), indicating possible over-correction by reweighting, while the drop of w/o Feature-Aware to 8.76 still confirms the necessity of feature modulation.

\section{Conclusion}
This letter proposes FADRW to address extreme class imbalance in few-shot linguistic steganalysis. Combining Dynamic Reweighting and Feature-Aware Modulation, FADRW mitigates decision bias and feature marginalization. Experiments on three real-world datasets show its superiority over existing methods, especially under 1\% steganographic ratios.

\clearpage
\bibliographystyle{IEEEtran}
\bibliography{refs}

\end{document}